\documentclass[12pt,preprint]{aastex}

\slugcomment{Accepted for publication in ApJ}

\shorttitle{On the Origin of Bimodal HBs in Massive GCs}

\shortauthors{Yoon et al.}

\begin{document}

\title{ON THE ORIGIN OF BIMODAL HORIZONTAL-BRANCHES IN MASSIVE GLOBULAR CLUSTERS:
\\THE CASE OF NGC 6388 and NGC 6441}

\author{Suk-Jin Yoon, Seok-Joo Joo, Chang H. Ree, Sang-Il Han, Do-Gyun Kim, and Young-Wook Lee}

\affil{Center for Space Astrophysics \& Department of Astronomy, Yonsei
University, Seoul 120-749, Korea}

\email{sjyoon@galaxy.yonsei.ac.kr}


\begin{abstract}

Despite the efforts of the past decade, the origin of the bimodal
horizontal-branch (HB) found in some globular clusters (GCs) remains
a conundrum. Inspired by the discovery of multiple stellar
populations in the {\it most massive} Galactic GC, $\omega$
Centauri, we investigate the possibility that two distinct
populations may coexist and are responsible for the bimodal HBs in
the {\it third} and {\it fifth} brightest GCs, NGC 6388 and NGC
6441. Using the population synthesis technique, we examine two
different chemical ``self-enrichment'' hypotheses in which a
primordial GC was sufficiently massive to contain two or more
distinct populations as suggested by the populations found in
$\omega$ Cen: (1) the age-metallicity relation scenario in which two
populations with different metallicity and age coexist, following an
internal age-metallicity relation, and (2) the super-helium-rich
scenario in which GCs contain a certain fraction of helium-enhanced stars, 
for instance, the second generation stars formed from the helium-enriched ejecta of the
first. The comparative study indicates that the detailed
color-magnitude diagram morphologies and the properties of the RR
Lyrae variables in NGC 6388 and NGC 6441 support the latter
scenario; i.e.,  the model which assumes a minor fraction ($\sim$ 15
\%) of helium-excess (Y $\simeq$ 0.3) stars.
The results suggest that helium content is the main driver behind
the HB bimodality found most often in massive GCs. If confirmed, the
GC-to-GC variation of helium abundance should be considered a {\it
local} effect, further supporting the argument that age is the {\it
global} second parameter of HB morphology.
\end{abstract}


\section{INTRODUCTION}

The horizontal-branch (HB) morphology of a globular cluster (GC) is defined as
the color distribution of its HB stars. The physical cause of the wide
diversity in the HB morphology among GCs with similar metallicity, or the
``second parameter effect", has long been the subject of examination
\citep{lee94,ste96,sar97,lee99a,bel01,cat05,rec06}. Although several other
variables may simultaneously play roles in HB morphology, the idea that GC age
is the major second parameter has been very popular \citep{lee94,rey01,yoo02}.
However, one of the strongest arguments against the age hypothesis has been the
existence of a number of GCs whose HB color distribution is distinctly bimodal
\citep{roo93,ste96}. An explanation based purely upon age as the second
parameter suggests that, at a given metallicity, the population associated with
the red HB is a few gigayears younger than that associated with the blue HB.
This cannot be reconciled with the traditional ``single-population" picture of
GCs.

Perhaps the most striking examples of HB bimodality are the two Galactic bulge
GCs, NGC 6388 and NGC 6441, along with NGC 2808 \citep{pio02,sos97}. NGC 6388
and NGC 6441 are relatively metal-rich GCs with [Fe/H] = $-$0.60 and $-$0.53,
respectively \citep{arm98}. Observations from the {\it Hubble Space Telescope}
({\it HST}\,) have revealed the presence of a significant fraction of extended
blue-HB stars \citep{pio97,ric97,pri03,pio02,bus04,cat06}, in addition to a
majority of red-HB stars as would be expected given their metallicity.

Recent studies suggest that the long-standing puzzle of the HB bimodality may
no longer be a complete mystery. It has been discovered that at least four
discrete populations coexist in the most massive Galactic GC, $\omega$ Cen
\citep[and references therein]{lee99b,sta06}. This finding obviously opposes
the conventional ``single-population'' picture of GCs, and also provides an
instructive precedent for the bimodal-HB feature in other GCs. The multiple populations
in $\omega$ Cen imply an internal age-metallicity relation (AMR), in that stars
having a higher metallicity are younger. Furthermore, subsequent observations
have revealed the presence of a marked double main-sequence (MS) with a
minority population of bluer and/or fainter MS stars, which are separated from
the majority \citep{and02,bed04}. \citet{nor04} and \citet{pio05} have
suggested that a greatly enhanced helium abundance ($\Delta$Y = 0.12 $\sim$
0.14) can explain the split MS of $\omega$ Cen. Lee et al. (2005) have further
shown that the helium-enhanced population in the MS is most likely the
progenitor of the extreme HB (EHB) stars.

It has also come to our attention that the integrated luminosities of NGC 6388
and NGC 6441 \citep{har96} make them {\it the third and fifth brightest} GCs
next to $\omega$ Cen and M54 among the $\sim$ 150 Galactic GCs
\citep{yoo00,ree02}. We are particularly inspired by the fact that five of the
10 brightest GCs can be characterized by the ``composite'' nature of
color-magnitude diagrams (CMDs), such as multiple red-giant-branches (RGBs) and/or
bimodal HBs \citep{van96}. The five GCs in this group are $\omega$ Cen, M54
\citep{sie07}, the two GCs of interest in this study, NGC 6388 and NGC 6441,
and the seventh brightest GC, NGC 2808 (Table 1, 6th column). On the other
hand, according to the \citet{har96} catalog, the 11 confirmed or suspected
bimodal-HB GCs \citep{cat98} are always brighter than the turnover of the
$V$-band luminosity function ($M_V^{tot}$ $\simeq$ $-$7) of the Galactic GC
system (Fig. 1$a$). The relative fraction of the combined sample of the
bimodal-HB GCs, $\omega$ Cen, and M54 progressively increases at the higher
brightness bin (Fig. 1$b$). It may be that there is a link between the massive
bimodal-HB GCs and the most massive multiple-population GCs -- $\omega$ Cen and
M54.

The purpose of this paper is to explore, with the stellar population synthesis
technique, the possibility that two distinct stellar populations may coexist
and be responsible for the HB bimodality in NGC 6388 and NGC 6441. We assume
the two chemical ``self-enrichment" processes that were proposed to explain
discrete subpopulations in $\omega$ Cen: (1) the two hypothetical
subpopulations in NGC 6388 and NGC 6441 follow an internal AMR (hereafter ``the
AMR scenario"), and (2) NGC 6388 and NGC 6441 contain a certain fraction of
super-helium-rich (SHR) stars (hereafter ``the SHR scenario''). Further
investigation is in progress to determine whether one of these scenarios can
account for other GCs thought to contain bimodal HBs \citep{cat98}, including
NGC 1851, NGC 2808, and NGC 6229.

The paper is organized as follows. A brief description of the stellar
population models is given in \S\,2. We compare the synthetic CMDs (\S\,3) and
the model RR Lyraes (\S\,4) to observed data. We argue that the bimodality of
HBs and the unusual properties of the RR Lyrae variables observed in NGC 6388
and NGC 6441 are better explained by the SHR scenario. Finally, we discuss the
implications of our results in \S\,5.

\section{STELLAR POPULATION MODELS}

The present models are constructed using the Yonsei Evolutionary Population
Synthesis (YEPS) code \citep{par97,lee99b,lee00,rey01,yoo02,lee05,yoo06}. The models
use a set of stellar libraries for the MS--RGB and post-RGB evolutionary tracks
that were built using the same input physics for consistency. 
Combined with RR Lyrae pulsation theories, the synthetic
population models can predict the properties of the RR Lyrae variables
\citep{lee90,yoo02}. It is not our intention to discuss the population
synthesis technique in this paper, and readers are referred to the above papers
and Yoon et al. (2008, {\it in prep.}) for the details of the YEPS model. Table
2 and Table 3 summarize the model's ingredients and input parameters employed
in this study.

\section{COLOR-MAGNITUDE DIAGRAMS}  

\subsection{The Age-Metallicity Relation (AMR) Model}  

Figure 2 compares our population models to the observed {\it HST} CMDs of NGC
6388 and NGC 6441 \citep{ric97}. We note that more recent {\it HST} CMDs now
exist for the GCs \citep{pri03,cat06}. For NGC 6441, however, we selected 
Rich et al.'s CMD instead of that of \citet{pri03} as the former is better
fit for the purposes of our CMD analysis in this section. The snapshot study by
Pritzl et al. is abundant in information on variable stars and clearly more
appropriate for the RR Lyrae analysis, so we fully exploit their results in
\S\,4. For the sake of consistency with the case of NGC 6441, we also opt for
Rich et al.'s CMD over the superseding CMD by \citet{cat06} for NGC 6388.

To simulate CMDs based on the AMR hypothesis (Figs. 2$b$ \& 2$e$), we first
generate model CMDs assuming a common age of 13 Gyr, but different [Fe/H] of
$-$0.60 for NGC 6388 and $-$0.53 for NGC 6441. Once the model shows reasonable
agreement with the MS through the RGB and {\it red} portions of the HB, we
then, according to our working hypothesis, add the second minor component
following a mild internal AMR. We adjust the AMR parameters until the model
CMDs mimic both the overall appearance and the population number ratios of the
HB. Given the level of photometric accuracy, the purpose of our CMD fit is
neither to determine the absolute age of the GCs nor to make a definitive
statement about their exact AMR parameters. For a direct comparison with
observations, we carried out Monte Carlo error simulations based on the actual
observational uncertainties.

The occurrence of a rather strong differential reddening across the fields of
NGC 6388 and NGC 6441 is a well-known phenomenon
\citep{pio97,hei99,lay99,rai02,law03}. Interestingly, the differential
reddening effect facilitates the development of markedly sloped red clumps. By
including such an effect, our model can reproduce the tilted red clump found in
the GCs. However, several studies have argued that the observed
differential reddening would not be enough to turn a ``normal'' red HB into a
significantly sloped structure \citep{cat96,bea01,rai02}. It is thus important
to assess the extent to which the models are able to reproduce the red clumps.

Figure 3 illustrates our numerical experiments on the impact of differential
reddening upon these structures. The middle panels (Figs. 3$b$ \& 3$e$) show
the flat red clumps without differential reddening, and the right panels (Figs.
3$c$ \& 3$f$) show the tilt of the clumps due to this effect. The simulations
suggest that a differential reddening value of $\sigma_{E(B-V)}$ = 0.03 (mag.)
is sufficient for the models to reproduce the observed tilt of the red clumps.
This is comparable to the observed value of $\Delta E(B-V)$ $<$ 0.1 ($\sim$ 3
$\times$ $\sigma_{E(B-V)}$) in \citet{rai02}, and slightly smaller than 
$\Delta$$E(B-V)$ $\simeq$ 0.15 ($\sim$ 5 $\times$ $\sigma_{E(B-V)}$) in 
\citet{pio97}, \citet{hei99}, \citet{lay99}, and \citet{law03}. 
Figure 3 also shows that the
models reproduce the observed RGB bump positions. It should be stressed that,
in addition to the shape of red HB clumps, the differential reddening effect seems
to be necessary to reproduce the observed scatter and possible tilt of the RGB
bumps.

We now turn back to Figure 2. The comparison of the CMDs between observed data
(Figs. 2$a$ \& 2$d$) and the models (Figs. 2$b$ \& 2$e$) shows reasonable
matches from the MS through the HB. The colors of RGB stars are sensitive to
metallicity, especially in the high-[Fe/H] regime. Yet, the metallicity
differences between the two hypothetical populations are small enough to
produce apparently single RGBs within observational uncertainties. Decreasing
age shifts the color of the RGB towards blue, and this effect, albeit small,
also helps to generate the single RGBs.

Another important result illustrated in Figure 2 is that the observed bimodal
HBs can be reproduced by the sum of the red HB (from the metal-rich younger
component) and the blue HB (from the metal-poor older component).
Interestingly, theories of stellar evolution predict that increasing age or
decreasing metallicity yields a bluer HB, whereas decreasing age or increasing
metallicity produces a redder HB \citep{lee94}. Hence, it is obvious that an
internal AMR splits readily the HB of each population in two, leading to bimodal HBs.
Moreover, our population models \citep{lee99a,rey01,yoo02} show that the HB
morphology is more sensitive to GC age by a factor of two when compared to
earlier studies \citep{lee94}. This results in a substantial reduction in the
required age difference between blue- and red-HB populations. This is significant
because all of the bimodal-HB GCs found to date \citep{cat98} have no
noticeable structure in the CMDs near the MS turn-offs \citep{pio02}, thus
leaving little room for an age spread of more than a couple of gigayears. The
difference in [Fe/H] and the ages required to reproduce their CMD morphologies
are 0.35 dex and 2.0 Gyr for NGC 6388, and 0.22 dex and 2.0 Gyr for NGC 6441
(Table 3).

There are several versions of the semi-empirical mass-loss formula
used for reproducing the observed amount of the mass loss in giant
stars. Various mass-loss formulae are summarized by \citet{cat00,cat05}. 
We note that a different choice of formulae
somewhat alters the required difference in [Fe/H] and age
between the blue and red HBs, but does not affect our conclusion.
For instance, according to Figure 11 of \citet{cat00}, the
\citet{rei75} formula would allow $\Delta$age = 2.0 Gyr centered at
13 Gyr to yield a 5 \% difference in $\Delta M$ (i.e., the mass loss
along the RGB). If one adopts Equation A5 (the steepest one) in
\citet{cat00}, then 5 \% in $\Delta M$ translates into $\Delta$age =
2.3 Gyr centered at 13 Gyr. Thus, even the most extreme choice of
the mass-loss formula would cause only a $\sim$ 15 \% [$=$
(2.3$-$2.0)$/$2.0] change in the age difference between red- and
blue-HB populations. In order to inspect the metallicity dependence
of the formulae, we take the steepest example in Figure 4 of
\citet{cat05} instead of the Reimers' formula, and find that the
required difference in [Fe/H] between red- and blue-HB populations
in the AMR model becomes smaller by about a factor of eight. This
makes the reproduction of bimodal HBs even easier.

The above analyses lead us to conclude that a mild internal AMR between the two
hypothetical populations may be a plausible cause of the HB bimodality,
together with the single MSs and RGBs within observational uncertainties.

\subsection{The Super-Helium-Rich (SHR) Model}  

We now consider the SHR hypothesis. The model CMDs in Figures 2$c$ \& 2$f$ are
constructed based on this assumption. To simulate the CMDs, we adopt [Fe/H] =
$-$0.60 for NGC 6388 and $-$0.53 for NGC 6441, with a common age of 13 Gyr and
a helium abundance of Y = 0.245. The major dominant populations are the same in
both the AMR (the metal-rich younger component denoted by ``1'' 
in Figs. 2$b$ \& 2$e$  ) and
the SHR (the normal-helium component denoted again by ``1'' 
in Fig. 2$c$ \& 2$f$) models.
We then add a helium-enhanced population according to our working hypothesis. The
helium abundance and the number of stars are adjusted until the best match
between the modeled and observed CMDs is achieved, especially for the blue HB
appearance. Observational errors and differential reddening effects are also
simulated in the same manner as in the AMR model.

\citet{pio05} report that the stars of the bluer component of the split MS in
$\omega$ Cen are more metal-rich (by 0.3 dex in [Fe/H]) than those of the
dominant redder component. For NGC 6388 and NGC 6441, however, there is no such
constraint dictated by observations. For a reference, in order for a 0.3 dex
more metal-rich population to mimic the observed blue-HB structure in NGC 6388
and NGC 6441, the required increase in helium abundance ($\Delta$Y) is 
as small as 0.02. Hence, in the
SHR scenario, we consider the helium-enriched population to have the same
[Fe/H] as the normal-Y population. The parameters required to reproduce
their CMD morphologies are listed in Table 3.

The model CMDs based on the SHR assumption match well with observations from
the MS through the HB. The variation in helium abundance has a relatively weak
effect on the MS to the RGB, so the simulations show no indication of
bifurcation on the MS through the RGB in each model CMD. The multiple MS feature,
as found in $\omega$ Cen \citep{pio05} and NGC 2808 \citep{pio07}, 
is not feasible in the available CMDs which contain
considerable observational uncertainty under the MS turnoffs. More importantly,
the observed bimodal HBs can be reproduced by the sum of the red HB (from the
normal helium component) and the blue HB (from the helium-enriched component).
For identical values of the total mass and helium core mass, theoretical models
show that the HB stars with a higher Y are bluer \citep{roo70,swe76}. Moreover,
for a given age and metal abundance, a helium-enriched HB star has a thinner
hydrogen envelope surrounding the helium-burning core, making the HB even
bluer. As a result, bimodal HBs with single RGBs can be achieved given
the assumption that $\sim$ 15 \% of the stars in NGC 6388 and NGC 6441 have an
enhanced helium abundance of $\Delta$Y $\simeq$ 0.05 (Table 3).

The CMD morphology analysis suggests that both the AMR and SHR models are
generally in first-order agreement with observations, reproducing the bimodal
HBs along with the single RGBs. We proceed to make a detailed
comparison of the HB structures in the two models.

\subsection{Comparison between the AMR and SHR Models}  

Apart from the overall agreement in the CMD morphology between the observations
and the models, there is an important feature of HBs that must be explained in
detail. The observed HBs seem to slope upward with decreasing $(B-V)$, and
their upper parts are brighter than the bulk of the red HBs by $\sim$ 0.5 mag
in $V$ \citep{ric97,swe98,rai02,pri03,cat06}.

The AMR models in Figures 2$b$ \& 2$e$ can only partially recreate the feature
by virtue of the HB evolutionary effect \citep{lee90}. If there are two HB
populations as predicted by the AMR model, the stars at the top of the blue HB
are most likely to be highly evolved and thus brighter than stars near the
zero-age HB (ZAHB). As a result, the HB evolutionary effect promotes the difference in
brightness between the upper part of the blue HB and the red clump. However,
inspection of Figures 2$a$ \& 2$d$ suggests that the RR Lyraes in NGC 6388 and
NGC 6441 appear to be dominated by stars near the ZAHB. It is therefore
unlikely that the brightness of the blue HBs of NGC 6388 and NGC 6441 support
the AMR scenario.

In contrast to the AMR model, the interesting HB appearance can be more readily
reproduced by the SHR model (Figs. 2$c$ \& 2$f$), because helium-rich HB stars
are intrinsically brighter. In an attempt to explain the sloping blue HB
extensions in NGC 6388 and NGC 6441, \citet{swe98} have first proposed a high
helium abundance (Y $\simeq$ 0.4). However, \citet{lay99} claimed that the
estimated Y values of NGC 6388 and NGC 6441 via the R-method strongly disfavor
that scenario. Layden et al. showed that the models in which Y = 0.38 and 0.43,
presented in Sweigart \& Catelan, inevitably yield R = 3.4 and 3.9. These are
about 3 $\sigma$ away from the observed value of R = 1.6 $\pm$ 0.7, which
corresponds to Y = $0.25^{+0.05}_{-0.08}$. Although our SHR model in this study
assumes a subpopulation with a high helium abundance (Y $\simeq$ 0.3), its
number fraction is as low as $\sim$ 15 \%. A calculation shows that the
number-weighted {\it mean} Y of the GCs in our SHR model is as low as Y
$\simeq$ 0.254 for NGC 6488 and 0.252 for NGC 6441, in complete accordance with
$0.25^{+0.05}_{-0.08}$ by \citet{lay99}.

Observations indicate that NGC 6388 and NGC 6441 contain a small portion of
additional EHB stars on their HBs. For the AMR model to reproduce this detail,
it should employ a unreasonably large mass dispersion ($\sigma_{M}$ $\sim$ 0.06 $M_{\odot}$) 
on the blue HBs. This is three times larger than 
the commonly used value of $\sigma_{M}$ = 0.02 $M_{\odot}$ \citep{lee94,cat98,yoo02},
which is currently used for the red HBs in our models.
In contrast, the SHR model requires only $\sigma_{M}$ $\simeq$ 0.03
$M_{\odot}$ to simulate the EHB structure. This is because, for larger helium
abundances, the blue loops in the evolutionary tracks can become considerably
longer, reaching higher effective temperatures \citep{swe87}.

Alternatively, there is a possibility of the presence of a minor
third subpopulation which is responsible for the EHBs in NGC 6388
and NGC 6441. In the AMR hypothesis, the third population presumably
has an older age than the underlying metal-poor old population. Our
model suggests an age as high as $\gtrsim$ 17 Gyr. Such an
unrealistically old age would not be acceptable. In the SHR model, 
on the other hand, the third population should have a larger $\Delta$Y 
than the blue-HB component. Our present model gives
a helium abundance of Y $\simeq$ 0.33 -- 0.34 for the EHB stars in
NGC 6388 and NGC 6441.

It is interesting to note that the SHR scenario is more consistent
with EHB (a. k. a. ``blue hook'') stars that are known to be fainter
than the redder HB stars and below the ZAHB locus.
\citet{lee05} show that the hottest EHB stars found in NGC 2808
\citep{bro01} are fainter than the redder HB stars as a natural
consequence of their high Y, consistent with \citet{swe87} predictions.
This may indicate that a high Y is sufficient {\it per se} to
explain the hottest blue-hook stars without necessarily invoking a delayed
helium flash \citep{lan04,moe04}. Readers are referred to
\citet{cat05} for various characteristics of EHB stars. The origin
of EHB stars and their possible link to the HB bimodality are
interesting issues worth further investigation. In this study we
consider the EHB stars as blue HB stars when calculating the number
ratios of blue and red HB stars.

\citet{cle05} have shown that the mean metallicity of the RR Lyrae stars in NGC
6441 is close enough to the typical metal abundance for this cluster.
More recently, \citet{gra07} have found no clear sign of
star-to-star Fe abundance variation in NGC 6441 from the Giraffe spectrograph at VLT2.
Hence, there appears to be a growing body of evidence that contradicts the
notion that RR Lyraes and blue HB stars have a lower metal abundance.
Moreover, \citet{moe06} have reported that the physical parameters of the cool blue HB
stars in NGC 6388 are consistent with the predictions of the helium enrichment
scenario, adding support to the SHR hypothesis.

We have compared the AMR and SHR models in terms of the CMD morphology, and
concluded that the SHR model is more successful at reproducing the details of
the observed CMDs of NGC 6388 and NGC 6441. There is yet another important
feature to be accounted for -- their unusual RR Lyrae properties. This is the
subject of the following section.

\section{RR LYRAE VARIABLE STARS} 

Observations \citep{swe98,lay99,cle01,pri00,pri01,pri02,pri03,cor06} have
revealed unusual properties of the RR Lyrae variables in NGC 6388
and NGC 6441. First, the mean pulsation periods of both the ${ab}$-
and $c$-type variables are too long for their metallicities, so they
do not fall into either of the usual Oosterhoff groups
\citep{oos39,yoo02}. Second, the number fraction of type $c$
variables, $N(c)/N(ab+c)$, falls between the two Oosterhoff groups,
but closer to that of the group II than of the group I. This is unusual because
the traditional group II GCs are always metal-poor ([Fe/H] $\leq$ $-$1.6).
These findings have posed yet another serious challenge to our
understanding of the GC stellar population. It is interesting to see
whether the present models based on the AMR and the SHR hypotheses
reproduce the peculiar properties of RR Lyrae variables in NGC 6388
and NGC 6441.

\subsection{The Pulsating Periods} 

Figure 4 presents a direct comparison between the observed and predicted RR
Lyraes in NGC 6388 and NGC 6441 in terms of their pulsating periods as a function of
$(B-V)_0$. In Figures 4$a$ \& 4$d$, we show 12 RR Lyraes found in NGC 6388
\citep{pri02}, and 63 RR Lyraes in NGC 6441 \citep{pri01,pri03}. To obtain the
fundamental periods ($P_F$), the $c$-type period ($P_c$) has been
fundamentalized using the equation, Log $P_F$ = Log $P_c$ + 0.13 \citep{cas87,lee90}. As
shown in Figure 2, the RR Lyrae stars are predicted to belong to the blue-HB
components in both the AMR and SHR models. Since the inferred physical
parameters of the blue-HB populations of NGC 6388 and NGC 6441 are almost
identical (Table 3), one may consider the RR Lyrae populations of the two GCs
to be twins. Indeed, Figure 4$a$ reveals that, despite a difference in the
observed number of RR Lyraes in NGC 6388 and NGC 6441, their RR Lyrae
distributions in the $P_F$ vs. $(B-V)_0$ diagram are statistically
indistinguishable from each other. On these grounds, we combine the catalogs of
NGC 6388 (12 RR Lyraes) and NGC 6441 (63 RR Lyraes) in order to minimize the
small number statistics.

The AMR model predicts shorter periods at a given $(B-V)_0$ than observations
show. Figures 4$b$ \& 4$e$ show examples of the Monte Carlo simulations. The
10,000 Monte Carlo realizations under the AMR assumption give $\langle P_{F}
\rangle$ = 0.555 $\pm$ 0.041 (day) from 12 $\pm$ 3 model RR Lyraes (NGC 6388) and
0.567 $\pm$ 0.017 (day) from 63 $\pm$ 7 (NGC 6441). Each of the $\langle P_{F}
\rangle$ values is 2.6 $\sigma$ and 5.5 $\sigma$ from the observed value of
$\langle P_{F} \rangle$ = 0.661 (day) for the combined NGC 6388 and NGC 6441 data.
For the AMR model to achieve a longer mean period, more highly-evolved RR
Lyraes and thus a bluer HB are required. However, the number of these evolved
and bright RR Lyrae stars is expected to decrease sharply as the HB morphology
gets bluer. This problem has been investigated in detail by \citet{pri02}, who
concluded that the model, which produces the observed ratio between blue HB
stars and RR Lyraes, cannot reproduce the exceedingly long mean period of NGC 6388.
Therefore, the AMR model appears to be in conflict with the available RR Lyrae
observations.

On the contrary, the SHR model appears to succeed in reproducing both the mean
$P_F$ and the observed $(B-V)_0$--$P_F$ correlations. Figures 4$c$ \& 4$f$ show
examples of the Monte Carlo simulations. The 10,000 Monte Carlo realizations
performed under the SHR assumption give $\langle P_{F} \rangle$ = 0.629 $\pm$
0.041 (day) from 12 $\pm$ 3 model RR Lyraes (NGC 6388) and 0.645 $\pm$ 0.018 (day) 
from 63 $\pm$ 7 (NGC 6441). Each of the $\langle P_{F} \rangle$ values exhibits 
$<$ 1 $\sigma$ agreement with the observed value 
of $\langle P_{F} \rangle$ = 0.661 (day).
As in the AMR model, all of the RR Lyraes belong to the blue-HB population in
the SHR model. However, this model does not require the presence of highly
evolved RR Lyraes to produce long-period RR Lyraes, as helium-enhanced RR
Lyraes are intrinsically brighter and of longer periods. Since most of the RR
Lyraes exist at the stage of the ZAHB, their significant numbers are also
reproduced. We emphasize that $\Delta$Y = 0.05 reproduces simultaneously the
peculiar RR Lyrae periods and the ratio between blue HB and RR Lyrae stars.

\citet{pri02} has attributed great importance to the presence of long-period
$c$-type RR Lyraes in NGC 6388 and NGC 6441, which are rarely found in other
GCs. The mean periods of $c$-type variables, $\langle P_c \rangle$, in NGC 6388
and NGC 6441 are 0.36 (day) and 0.38 (day). Note that the average $\langle P_c
\rangle$ for Galactic GCs is known to be 0.33 (day), respectively. In the AMR
model, the assumed $c$-types (i.e., $(B-V)_0$ $\lesssim$ 0.3) are 
concentrated toward the ZAHB (Figs. 4$b$ \& 4$e$) and thus have short periods. 
To include evolved $c$-types, the blue HB component must become bluer.
However, the model shows that this shift gives rise to a significant decrease
in the number of RR Lyraes, as discussed by \citet{pri02}. In contrast to the
AMR hypothesis, a raise in $\langle P_c \rangle$ agrees with the SHR scenario.
The high helium content increases the zero-point of RR Lyrae brightness,
leading to $c$-types of longer periods (Figs. 4$c$ \& 4$f$).

\subsection{The Oosterhoff Groups} 

Figure 5 shows the correlation between the mean period of type {\it ab} RR
Lyraes ($\langle P_{ab} \rangle$) in GCs and their [Fe/H]. In Figure 5$a$, we
display the well-known Oosterhoff dichotomy \citep{oos39}, along with the
unusual distribution of NGC 6388 and NGC 6441. The $\langle P_{ab} \rangle$
data are obtained from \citet{cle01} and \citet{cor06}. One can see that the
Galactic GCs are mainly divided into two distinct groups according to mean
period and metal abundance: Oosterhoff group I (filled squares; $\langle P_{ab}
\rangle$ $\simeq$ 0.55 day, and [Fe/H] $>$ $-$1.8) and group II (open squares;
$\langle P_{ab} \rangle$ $\simeq$ 0.65 day, and [Fe/H] $<$ $-$1.6). NGC 6388
and NGC 6441 (solid diamonds) have $\langle P_{ab} \rangle$ values of 0.676
(nine $ab$-types) and 0.759 (42 $ab$-types), respectively. Note that NGC 6388
and NGC 6441 belong to neither Oosterhoff group. It is \citet{pri00} who first
showed that NGC 6388 and NGC 6441 do not fall into either of the usual
Oosterhoff groups, and proposed the possibility of a new Oosterhoff class. Do
NGC 6388 and NGC 6441 break down the traditional Oosterhoff classification
scheme? Or, do we need to add a third class to the scheme?

The answers to these questions depend solely on our understanding of the origin
of the Oosterhoff dichotomy. An explanation for the phenomenon
has been put forward in the work of \citet{lz90} and \citet{yoo02}. In
particular, \citet{yoo02} have discovered that most of the lowest-metallicity
([Fe/H] $<$ $-$2.0) GCs display a striking planar alignment in the outer-halo of 
the Milky Way. The alignment, combined with evidence from kinematics and stellar
population, indicates that the metal-poorest GCs were originated  
from a satellite system, very likely from the Large Magellanic Cloud
(LMC). In such a case, age and metallicity can be decoupled, so the
lowest-metallicity Galactic GCs are not necessarily the oldest component of the
Milky Way. This is interesting because the Oosterhoff dichotomy among the
Galactic GCs can be naturally reproduced by assuming that the
lowest-metallicity GCs are slightly younger than the genuine Galactic GCs of
similar metallicity.

Figure 5$b$ conveys the essence of the \citet{yoo02} explanation. The Galactic
GCs are re-classified according to their Galactocentric distances ($R_G$), and
shown along with the population models. From this figure, one can see that the
inner- ($R_G$ $<$ 8 kpc) and outer-halo ($R_G$ $>$ 8 kpc) GCs are well
represented by the models of the 13-Gyr population (solid line), and by those
of a slightly younger population (11.8 Gyr; dashed line), respectively. The
observed $\langle P_{ab} \rangle$ of the lowest-metallicity ([Fe/H] $<$ $-$2.0)
GCs are well reproduced by the dashed line. This line also fits the GCs on
retrograde orbits, which are most likely an accreted component of the Milky Way
\citep{zin93,van93}. Note also that the LMC GCs follow the same dashed line.
From this analysis, \citet{yoo02} conclude that the Oosterhoff dichotomy can be
viewed as a manifestation of the age structure of the Galactic GC system, and
thus, of the Galactic assembly history.

Following the \citet{yoo02} framework, we now compare the AMR and SHR models
for NGC 6388 and NGC 6441 in terms of the $\langle P_{ab} \rangle$ vs. [Fe/H]
correlation (Fig. 5$b$). The use of the input parameters in Table 3 leads the
AMR model to predict an $\langle P_{ab} \rangle$ that is too low at their
[Fe/H] (the lower stripe in Fig. 5$b$). This result is in line with
\citet{pri02} and \citet{cat03}, who confirmed the difficulty in reproducing
the RR Lyrae periods for the given metallicities of NGC 6388 and NGC 6441. We
have explored a large parameter space to find a way to reproduce the
observations of NGC 6388 and NGC 6441 based on the AMR assumption, but failed
unless we assumed an unreasonably high age ($>$ 17 Gyr). Such an
unrealistically old age is not acceptable. Moreover, the very old population is
predicted to have too few RR Lyrae stars when compared with the observations.

By contrast, the SHR model is remarkably consistent with the observations
displayed in the Oosterhoff diagram (the upper stripe in Fig. 5$b$). We recall
that all of the RR Lyraes belong to the blue-HB population in both the AMR and
SHR models. When assuming that the helium abundance of the RR Lyraes is
enhanced by $\Delta$Y = 0.05 (Table 3), their large $\langle P_{ab} \rangle$
values can be easily achieved at their [Fe/H], even with RR Lyraes near the
ZAHB. This is not surprising, since any effect that leads to a brighter
HB such as an enhanced Y, would lead to longer periods, more consistent with
the observations \citep{swe98,swe99}.

We finish this section by discussing the unusual values of
$N(c)/N(ab+c)$ of NGC 6388 and NGC 6441 for their metallicities. 
The combined data of NGC 6388 and NGC 6441 give 0.40 (= 34/85) \citep{cor06}.
The value falls between Oosterhoff group I ($\sim$ 0.2) and group II ($\sim$ 0.5).
The $c$-type fraction of RR Lyrae variables has long been known
to vary among GCs with differing HB types \citep{lee90}. 
The $c$-type fraction is significantly higher for predominantly
blue HBs with little ZAHB stars within the RR Lyrae strip
(Oosterhoff group II) than for red or intermediate HBs (group I). 
\citet{lee90} demonstrated that the evolutionary effect of blue HBs causes an
uneven distribution in $(B-V)$, since the speed of evolution
increases as a star evolves further from its original position. As a
result, the GCs with predominantly blue HB components are predicted to
have high incidences of bluer RR Lyraes, i.e., $c$-type variables.
In this regard, the GCs with 0.3 $\lesssim$ $N(c)/N(ab+c)$ $\lesssim$ 0.4 
can only be produced by a transitional HB morphology 
from intermediate-HB to predominantly-blue HBs. 
Under the normal-Y circumstances, GCs with such a transient HB type are rare,
because it is hard for blue HBs to contain ZAHBs long enough to extend into the RR Lyrae strip.
Few examples include NGC 5904 and NGC 6626 with $N(c)/N(ab+c)$ $\sim$ 0.3. 
By constrast, the SHR model can generate relatively easily
blue HBs with long ZAHBs extending into the instability strip. 
This is because the high-Y HB tracks are characterized by considerably stretched
blue loops \citep{swe87}. The unusual value of
$N(c)/N(ab+c)$ ($\sim$ 0.4) for NGC 6388 and NGC 6441, therefore, points toward 
the SHR scenario.

By means of the Oosterhoff class argument, we conclude that the SHR
hypothesis provides a more plausible solution to the problem posed by the
unusual behaviors of the RR Lyraes in NGC 6388 and NGC 6441. As suggested by
\citet{pri00}, NGC 6388 and NGC 6441 may be taken as a third class of
Oosterhoff groups, and the helium abundance of RR Lyrae stars is the main
driver behind this new class.

In this section, we have compared the two models in terms of RR Lyrae
properties (i.e., the pulsating periods and the Oosterhoff classes), and
consistently found that the observations are strongly in favor of the SHR
scenario over the AMR scenario.

\section{DISCUSSION}

Using stellar population simulations, we have investigated the possibility that
the bimodality of the HBs in NGC 6388 and NGC 6441 can be attributed to the
presence of two distinct populations. This study was motivated by the discovery
of multiple stellar populations coexisting in the most massive Galactic GC,
$\omega$ Cen \citep[and references therein]{lee99b,sta06}. We are particularly
inspired by the fact that the integrated luminosities of NGC 6388, NGC 6441,
and NGC 2808 \citep{har96} make them {\it the 3rd, 5th, and 7th} brightest
among the $\sim$ 150 Galactic GCs \citep{yoo00,ree02}.

Two chemical enrichment scenarios for GCs are discussed; namely, the
age-metallicity relation (AMR) and the super-helium-rich (SHR) scenarios. In
both models, two hypothetical populations coexist within individual GCs. We
have examined the CMD morphologies and the RR Lyrae properties of NGC 6388 and NGC
6441, and found consistently that the SHR scenario is superior to the AMR
hypothesis in accounting for the observations. Our best solution is
that the blue HB and the RR Lyrae stars are comprised of helium-enhanced
($\Delta$Y $\sim$ 0.05) stars, which make up $\sim$ 15 \% of the total
population. Table 4 summarizes the model's ability to reproduce
various aspects of the observations.

Recently, \citet[hereafter CD07]{cal07} have investigated the NGC 6441 HB population, 
including the periods of the RR Lyrae variables. 
From their analysis, CD07 conclude that Y
$\simeq$ 0.35 is required for the tilted red clump, Y $\simeq$ 0.37
for the RR Lyrae stars, and Y $\simeq$ 0.38 -- 0.40 for the blue HB.
A partial, qualitative
agreement is found between our SHR model and CD07's, in that the SHR
population is necessarily invoked. However, several factors
differentiate our result from CD07's: ({\it a}) While our model
assumes two discrete populations with distinct helium abundances,
CD07 present a continuous distribution of helium content; ({\it b})
The helium abundance required in our SHR model to reproduce the blue HB is Y $\simeq$
0.29 -- 0.30, which is well below the inferred Y in CD07 (Y $\simeq$
0.35 -- 0.40); ({\it c}) Our model exhibits the tilted red clump
naturally reproduced by the inclusion of differential reddening
effect, thus indicating the red clumps of a normal helium
abundance (Y $\simeq$ 0.24 -- 0.25). In short, the main difference 
lies in the predicted total helium abundance of the GCs,
in that our SHR model requires significantly
lower value of total helium abundance.
CD07 predict the total helium-enriched population amounting to $\sim$
60 \% of the entire stellar content. 
We note that, as mentioned in \S\S\,3.3, the R-method argument by \citet{lay99} 
allows only a limited amount of helium content in the GCs. 
Furthermore, using deep {\it HST}
photometry of NGC 6388, \citet{cat06} discover the
lack of a sizeable luminosity difference between the red HBs of NGC
6388 and 47 Tuc (a ``flat''-red-clump GC with similar metallicity).
This result points to the normal helium abundance of the red clump in NGC
6388, more consistent with our conclusion than CD07's.

The high-helium hypothesis was originally suggested to explain the
presence of hot EHBs by \citet{dan02}. This notion has been expanded
to account for the bimodal HBs found in NGC 2808
\citep{dan04,lee05}, NGC 6388, and NGC 6441 \citep[CD07, this
study]{dan04,dan05,dan06,cat05}. It is noteworthy that the GCs with
bimodal HBs often, if not always, also possess EHBs. 
Since the EHB phenomenon appears to share a common origin
with HB bimodality, it is important to see whether the occurrence of
EHBs correlates with GC mass, as does the instance of the bimodal
HB (see Figure 1 of \citet{lee07}). 
In fact, this was the basis for suggestions that the integrated luminosity
plays an important role in the context of the
second-parameter phenomenon \citep{fus93}. Table 1 shows that eight
Galactic GCs out of the 10 brightest contain hot EHB components (the
7th column). According to \citet{cat98} catalog, $\sim$ 90 \% (14
out of 16) of the EHB GCs have blue HBs. Given the
fact that increasing metallicity makes an HB redder, we speculate
that the EHB GCs with normal blue HBs in Table 1 (NGC 6266, NGC
7078, and NGC 6273) could have typical {\it bimodal} HBs if they
were as metal-rich as NGC 6388 and NGC 6441.

Several mechanisms have been proposed to explain the peculiar
chemical history of GCs with helium-excess populations
\citep{dan02,ven01,ven02,nor04,gra04,bek06,chu06,cho07,tsu07,new07}.
One theoretical requirement for the proposed chemical evolution is
an initially high mass to retain the ejecta of the first generation
stars, as is evident in the case of $\omega$ Cen
\citep{nor04,pio05,lee05,bek06}. The HB bimodality found in massive
GCs such as NGC 6388, NGC 6441 \citep[CD07, this
study]{dan04,dan05,dan06,cat05}, and NGC 2808
\citep{dan04,lee05,pio07} appears to be another piece of evidence
for a link between the GC mass and the helium-enrichment
process. It has been suggested that $\omega$ Cen was once part of a
more massive system that later merged with the Milky Way \citep[and
references therein]{lee99b,din02,alt05,mez05,bek06}. Similar
accretion events may have continued throughout the Galactic history,
and the massive GCs with bimodal HB distributions may be considered
minor versions of $\omega$ Cen, representing relics of the Galaxy
assembly process (see \citet{lee07} for further discussion).

Another important implication of our results concerns 
the long-standing second parameter debate regarding HB morphology. 
Given that the HB-bimodality phenomenon has long been taken 
as the strongest evidence against the GC age being the second parameter,
a fully satisfactory solution will not be found 
until the origin of the HB bimodality has been accounted for. 
Our results suggest that the GC-to-GC variation in helium abundance,
which was one of the first candidates for the
second parameter in the literature \citep{san67,van67},
have a role in giving rise to the HB bimodality. 
If our interpretation is confirmed,
the GC-to-GC helium variation found most often in massive GCs should be considered 
a {\it local} effect rather than a global one, 
further supporting age as the {\it global} second parameter.
Further observations and modelling of GCs with bimodal HBs are still required
to verify that helium variation is a local effect, 
and thus the third parameter controlling HB morphology.

\acknowledgments We would like to thank Michael Rich for providing the
observational data in Figures 2 \& 3. Helpful comments from the referee were
gratefully appreciated. 
S.-J. Y. acknowledges support 
from the Basic Research Program of the Korea Science \& Engineering Foundation 
(Grant No. R01-2006-000-10716-0), 
and from the Korea Research Foundation Grant (MOEHRD: KRF-2006-331-C00134).



\clearpage

\begin{deluxetable}{clcrllll}
\tablewidth{17cm}
\tabletypesize{\scriptsize}
\tablecaption{The 10 Brightest Galactic Globular Clusters and the Features in their CMDs\label{tbl-1}}
\tablehead{
\colhead{Rank} & \colhead{NGC} & \colhead{Name} & \colhead{$M_V^{tot}$ \tablenotemark{a}} &
\colhead{[Fe/H]\tablenotemark{b}} & \colhead{Composite CMD\tablenotemark{c} ?} &
\colhead{HB Shape} &\colhead{References}
}
\startdata
  1 &   5139 & $\omega$ Cen & $-$10.29 & $-$1.59 & Yes (Multiple MS/HB/RGB)	& {\bf Multiple HB} + EHB   & 1, 2, 3, 4   \\
  2 &   6715 & M54          & $-$10.01 & $-$1.43 & Yes (Multiple RGB/HB)   	& {\bf Multiple HB} + EHB   & 5, 6, 7   \\
  3 &   6388 & \nodata      &  $-$9.82 & $-$0.60 & Yes (Bimodal HB)     	& {\bf Bimodal HB} + EHB    & 8, 9, 10      \\
  4 &   2419 & \nodata      &  $-$9.58 & $-$2.10 & No               		& Normal blue HB            & 8         \\
  5 &   6441 & \nodata      &  $-$9.47 & $-$0.53 & Yes (Bimodal HB)     	& {\bf Bimodal HB} + EHB    & 8, 9, 11   \\
  6 &   104 & 47 Tuc        &  $-$9.42 & $-$0.71 & No                   	& Normal red HB             & 9         \\
  7 &   2808 & \nodata      &  $-$9.36 & $-$1.37 & Yes (Multiple MS/HB)     	& {\bf Bimodal HB} + EHB    & 9, 12, 13     \\
  8 &   6266 & M62          &  $-$9.19 & $-$1.29 & No                   	& Long blue HB + EHB        & 9         \\
  9 &   7078 & M15          &  $-$9.17 & $-$2.17 & No                   	& Normal blue HB + EHB      & 9         \\ 
 10 &   6273 & M19          &  $-$9.08 & $-$1.68 & No                   	& Normal blue HB + EHB      & 9         \\
\enddata
\tablenotetext{a}{$M_V^{tot}$ is obtained from \citet{har96}.}
\tablenotetext{b}{[Fe/H] is obtained from \citet{zin93}.}
\tablenotetext{c}{CMDs with multiple RGBs and/or HBs.}
\tablerefs{ (1) \citet{lee99b}; (2) \citet{hil00}; (3)
\citet{pan00}; (4) \citet{pio05}; (5) \citet{bel99}; (6) \citet{lay00}; 
(7) \citet{ros04}; (8) \citet{ric97}; (9) \citet{pio02}; 
(10) \citet{cat06}; (11) \citet{pri03}; (12) \citet{sos97}; (13) \citet{pio07} }
\end{deluxetable}

\clearpage

\begin{deluxetable}{ll}
\tabletypesize{\footnotesize}
\tablewidth{16cm}

\tablecaption{Model Ingredients\label{tbl-2}}
\tablehead{
\colhead{Ingredient} & \colhead{Stellar \& Flux Libraries} } \startdata
MS to RGB Evolutionary Tracks   \dotfill & \citet{kim02} Yonsei--Yale Normal-helium Isochrones                      \\
                     & Kim et al. (2008, {\it in prep.}) Yonsei--Yale Helium-rich Isochrones    \\
Post-RGB Evolutionary Tracks    \dotfill & Han et al. (2008, {\it in prep.}) Yonsei--Yale Normal-helium HB Tracks     \\
                     & Han et al. (2008, {\it in prep.}) Yonsei--Yale Helium-rich HB Tracks     \\
Flux Library  \dotfill & \citet{lej98} Model Atmosphere                                     \\
RR Lyrae Variable Stars   \dotfill  & \citet{lee90} Prescriptions                                   \\
\enddata
\end{deluxetable}

\clearpage

\begin{deluxetable}{lcc}
\tabletypesize{\footnotesize}
\tablewidth{14cm}

\tablecaption{Model Input Parameters\label{tbl-3}}

\tablewidth{0pt} 
\tablehead{ \colhead{Parameter} & \colhead{NGC 6388} & \colhead{NGC 6441}}
\startdata
\sidehead{{\bf Common Parameters:}} Initial mass function \dotfill & Salpeter       & Salpeter  \\
$\alpha$-element enhancement, $[$$\alpha$/Fe$]$  \dotfill       & 0.3                & 0.3          \\
Distance modulus, $(V-M_{V})$ (mag.)           \dotfill       & 16.50              & 17.17             \\
Galactic reddening, $E(B-V)$ (mag.)      \dotfill       & 0.34           & 0.43         \\
Differential reddening, $\sigma_{E(B-V)}$ (mag.)     \dotfill       & 0.03           & 0.03         \\
Extinction coefficient, $A_{V}/E(B-V)$       \dotfill       & 3.1                & 3.1           \\
HB mass dispersion, $\sigma_{M}$ ($M_{\odot}$)               \dotfill       & 0.02       & 0.02 \\
RR Lyrae instability strip width, $\Delta$\,Log\,$T_{e}$ (K)  & 0.085      & 0.085          \\
Reimers' (1975) mass-loss efficiency parameter, $\eta$ \dotfill         & 0.56      & 0.56    \nl

\sidehead{{\bf The AMR Model\tablenotemark{a}:}}
Metal abundance, Z \dotfill              & 0.007 \& 0.0035    & 0.008 \& 0.0055    \\
Helium abundance, Y \dotfill             & 0.245 \& 0.245      & 0.245 \& 0.245      \\
Absolute age, t (Gyr) \dotfill             & 13.0 \& 15.0      & 13.0 \& 15.0      \\
Number fraction (\%) \dotfill           & 82 \& 18      & 86 \& 14      \\

\sidehead{{\bf The SHR Model\tablenotemark{b}:}}
Metal abundance, Z \dotfill          & 0.007 \& 0.007    & 0.008 \& 0.008    \\
Helium abundance, Y \dotfill             & 0.245 \& 0.295      & 0.245 \& 0.295      \\
Absolute age, t (Gyr) \dotfill             & 13.0 \& 13.0      & 13.0 \& 13.0      \\
Number fraction (\%) \dotfill           & 82 \& 18      & 86 \& 14          \\

\enddata
\tablenotetext{a} {Metal-rich younger \& metal-poor older populations}
\tablenotetext{b} {Normal-helium \& helium-enriched populations}
\end{deluxetable}


\clearpage

\begin{deluxetable}{lcc}
\tabletypesize{\normalsize}
\tablewidth{14cm}

\tablecaption{Summary of Model Ability to Reproduce Observations\label{tbl-4}}

\tablewidth{0pt} 
\tablehead{ \colhead{Feature} & \colhead{       The AMR Model} & \colhead{The
SHR Model} } \startdata
Single RGB          \dotfill    & $\bigcirc$ & $\bigcirc$   \\
RGB Bump Luminosity            \dotfill    & $\bigcirc$ & $\bigcirc$   \\
RGB Bump Slope           \dotfill    & $\bigcirc$ & $\bigcirc$   \\
HB Bimodality       \dotfill    & $\bigcirc$ & $\bigcirc$   \\
Red HB Tilt         \dotfill    & $\bigcirc$ & $\bigcirc$   \\
Overall HB Slope    \dotfill    & $\times$   & $\bigcirc$   \\
Extreme HB           \dotfill    & $\times$   & $\triangle$  \\
RR Lyrae Luminosity \dotfill    & $\times$   & $\bigcirc$   \\
RR Lyrae Period     \dotfill    & $\times$   & $\bigcirc$   \\
$N(c)/N(ab+c)$      \dotfill    & $\bigcirc$ & $\bigcirc$   \\
Oosterhoff Diagram  \dotfill    & $\times$   & $\bigcirc$   \\
\enddata
\end{deluxetable}

\clearpage
\begin{center}
\includegraphics[width=12cm]{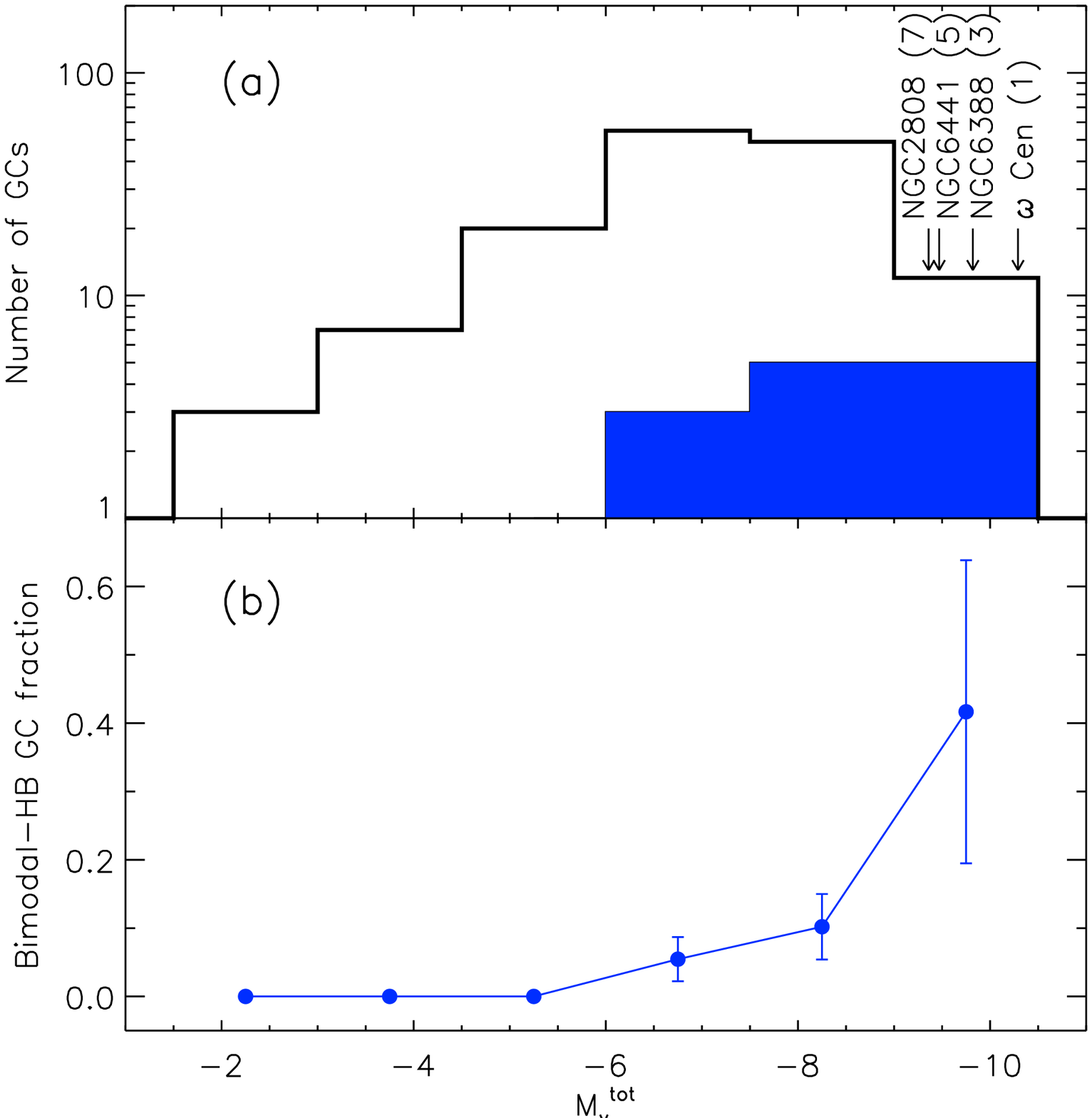}
\end{center}
\begin{figure}
\begin{center}
\caption{($a$) The $V$-band luminosity function (open histogram) of 147
Galactic GCs from \citet{har96} catalog. Arrows denote the integrated magnitude of the
brightest GC, $\omega$ Cen, and the GCs with bimodal HBs among the 10 brightest
GCs. The numbers in the parentheses denote their brightness ranks. NGC 6388 and
NGC 6441 are among the most massive GCs. The filled histogram is for the 11
bimodal-HB GCs \citep{cat98} plus $\omega$ Cen and M54. Note that
they are all brighter than $M_V^{tot}$ $\simeq$ $-$7. 
($b$) The relative number fraction of bimodal-HB GCs plus $\omega$ Cen and M54
as a function of the integrated magnitude. Their
fraction in a luminosity bin increases with increasing luminosity. Error bars
show Poisson errors. \label{fig1}}
\end{center}
\end{figure}

\clearpage
\begin{center}
\includegraphics[width=15cm]{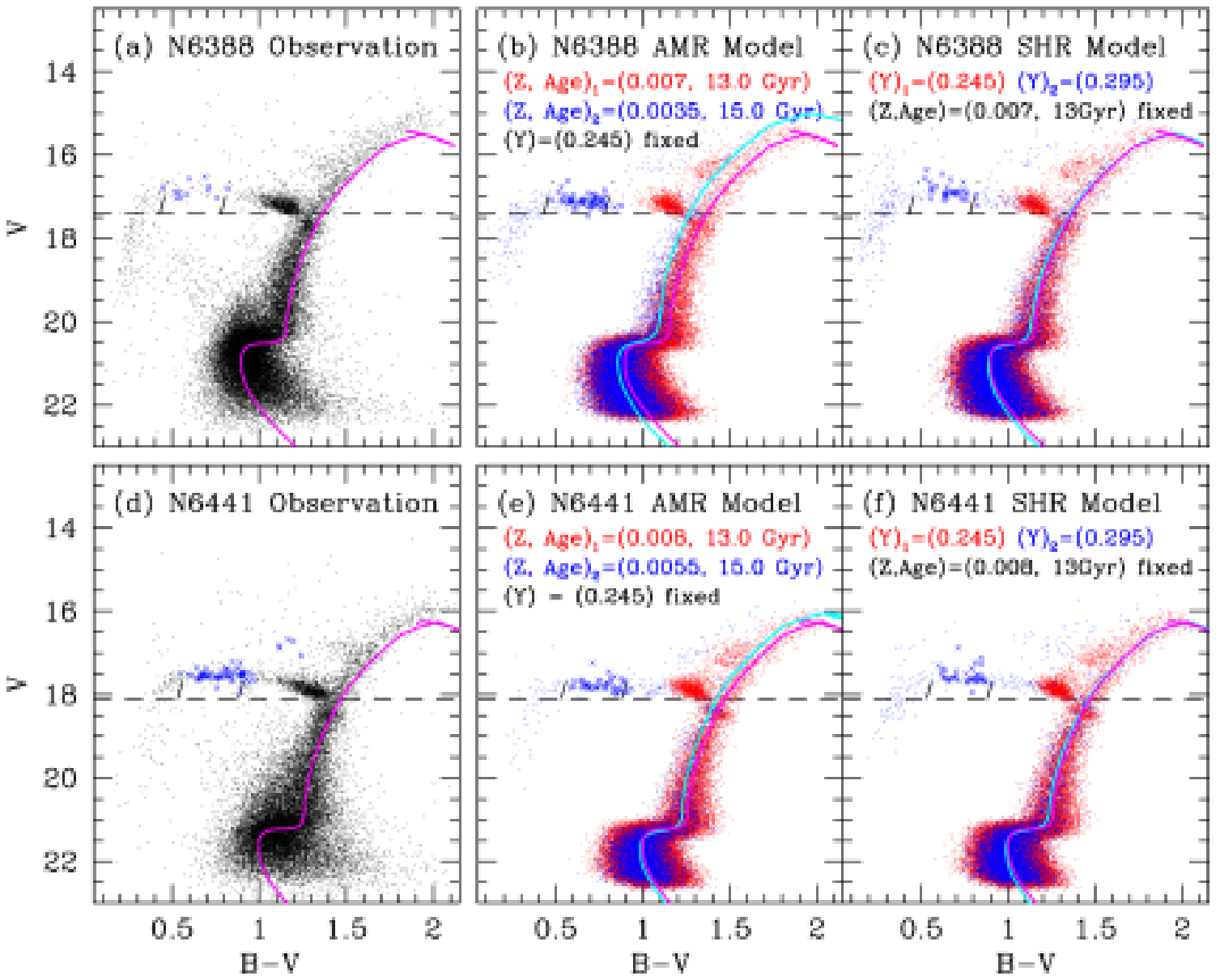}
\end{center}
\begin{figure*}
\begin{center}
\caption{Comparison of the observational data ($a$ \& $d$) with the AMR model ($b$
\& $e$) and the SHR model ($c$ \& $f$) for NGC 6388 and NGC 6441. $($$a$ \&
$d$$)$ The observed CMDs are from \citet{ric97}. Isochrones are overlaid with
observations, from which the dominant major populations, i.e., the metal-rich younger population in
the AMR model and the normal-helium population from the SHR model, are generated. 
Crosses represent the observed RR Lyrae stars from
\citet{pri01,pri02,pri03}. The same RR Lyrae stars are also shown in Fig. 4.
Horizontal dashed lines represent the brightness at the base of red clumps.
Short tilted solid lines denote the RR Lyrae instability strips. $($$b$, $c$,
$e$, \& $f$$)$ In $b$ \& $e$, the metal-rich younger (denoted by ``1'') and the
metal-poor older (denoted by ``2'') populations are respectively shown as red
and blue dots. The magenta and cyan isochrones are respectively for the
metal-rich younger and the metal-poor older populations. The combinations of
metallicity, helium abundance, and age required to reproduce the CMD morphology
are denoted. Note that the observed bimodal HBs are explained as the sum of the
red HBs from component 1 (red dots) and the blue HBs from component 2 (blue
dots). Crosses denote the model RR Lyrae stars. In $c$ \& $f$, the
normal-helium (denoted by ``1'') and the helium-excess (denoted by ``2'')
populations are respectively shown as red and blue dots. The magenta and cyan
isochrones are respectively for the normal-helium and the helium-excess
populations. Note that the observed bimodal HBs are explained 
as the sum of red HBs from component 1 (red dots) and
blue HBs from component 2 (blue dots). Crosses denote the model RR Lyrae stars.
\label{fig2}}
\end{center}
\end{figure*}

\clearpage
\begin{center}
\includegraphics[width=17cm]{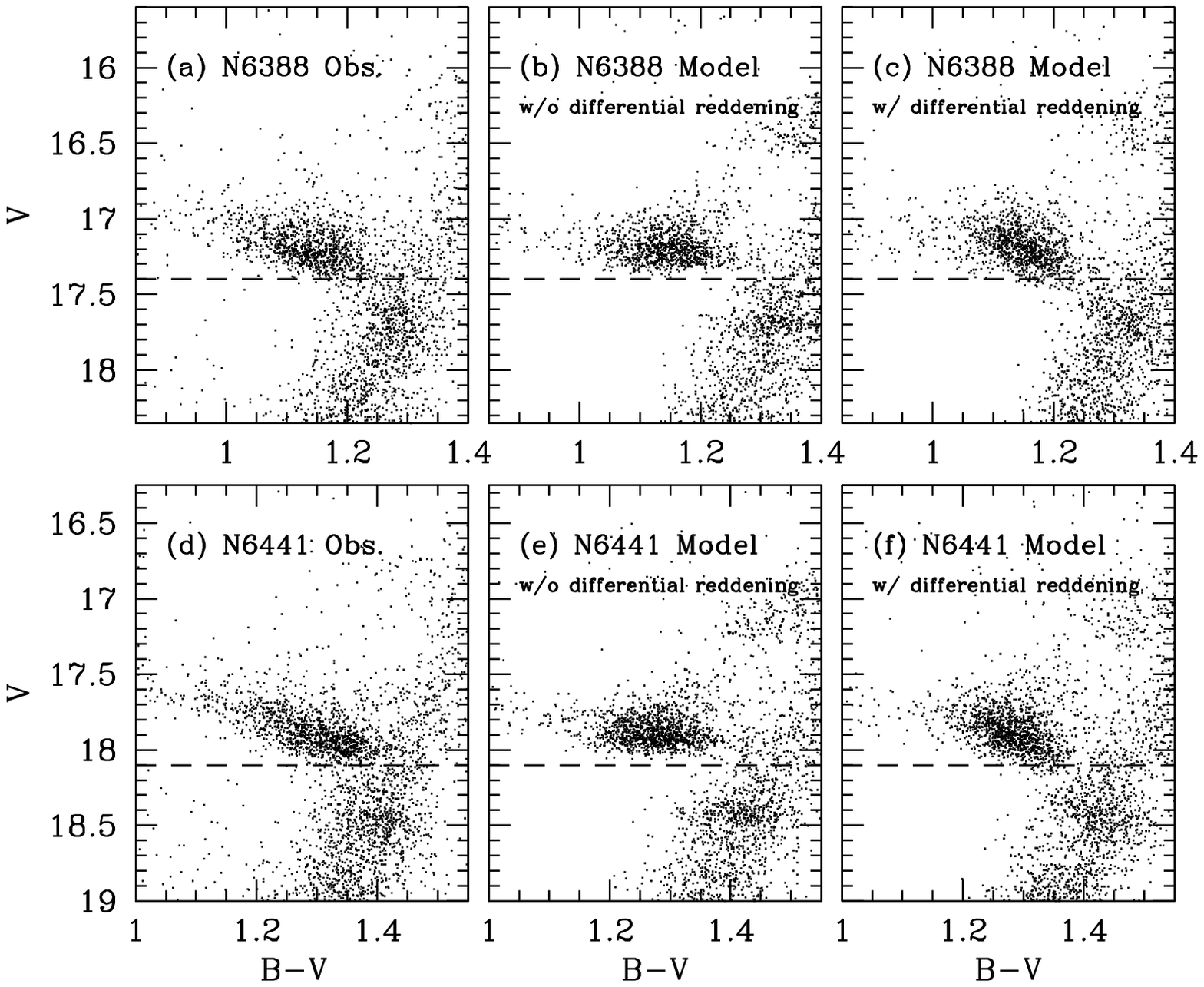}
\end{center}
\begin{figure*}
\begin{center}
\caption{Numerical experiments on the effect of differential reddening on the
red clumps. The observed clumps ($a$ \& $d$) are compared with the flat clumps
generated without the differential reddening effect ($b$ \& $e$), and the
tilted clumps generated with the effect ($c$ \& $f$). Note that panels ($a$ \&
$d$) and ($c$ \& $f$) are identical to Figs. 2$a$ \& 2$d$ and Figs. 2$c$ \& 2$f$,
but have been magnified around the HB red clump region. The
dominant major population is common across the AMR (the metal-rich younger
component denoted by ``1'' in Fig. 2) and SHR (the normal-helium component
denoted again by ``1'' in Fig. 2) models. The inferred differential reddening
appears to be enough to turn a ``normal'' red HB component into a significantly
sloped structure. Note also that the differential reddening effect appears to
be necessary to reproduce the observed scatter and the possible tilt of the RGB
bumps. \label{fig3}}
\end{center}
\end{figure*}

\clearpage
\begin{center}
\includegraphics[width=17cm]{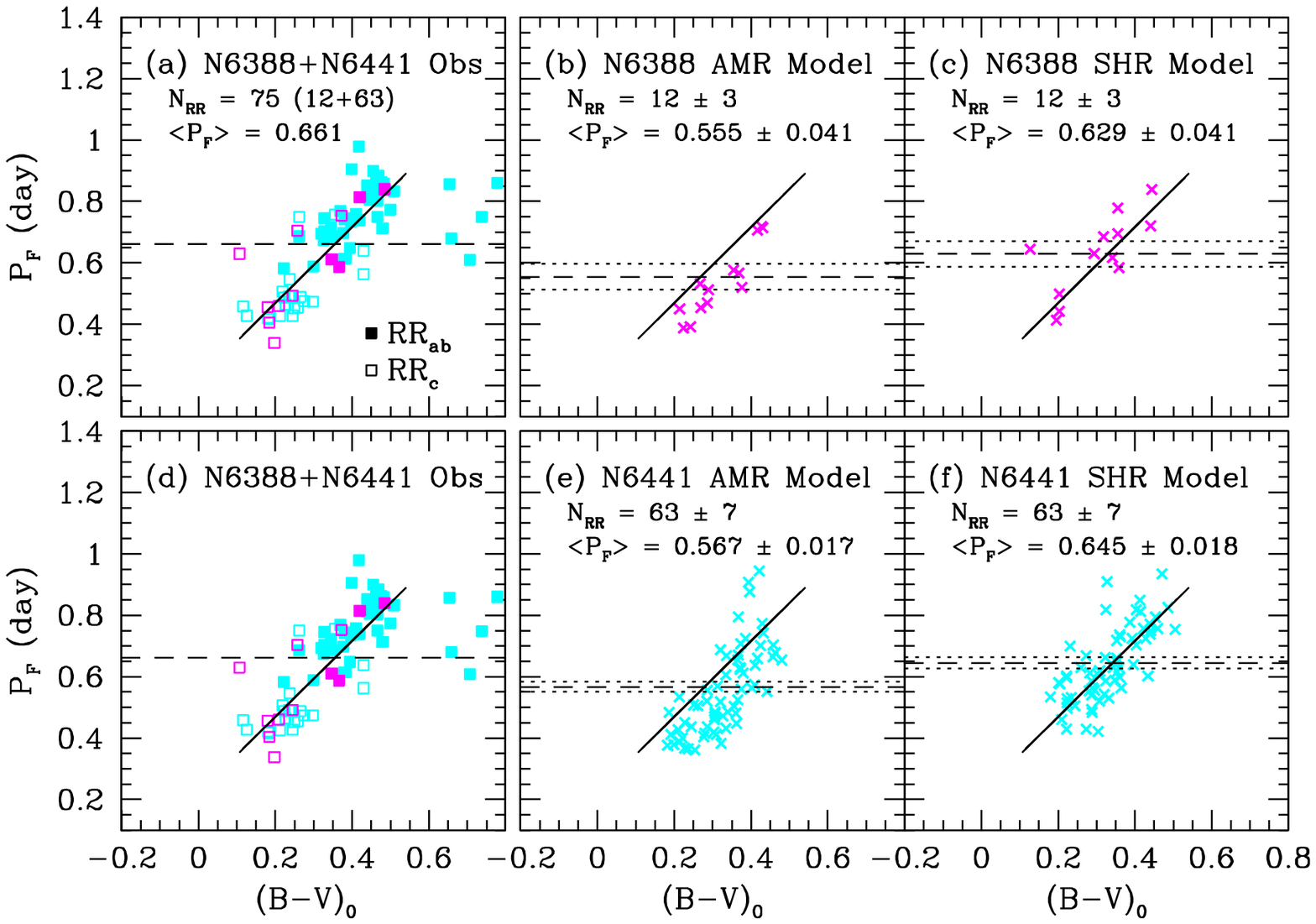}
\end{center}

\begin{figure*}
\begin{center}
\caption{Comparison of the observations ($a$ \& $d$) with the AMR model ($b$ \&
$e$) and the SHR model ($c$ \& $f$) for NGC 6388 and NGC 6441 in the ($B-V$,
$\langle P_{F} \rangle$) plane. $($$a$ \& $d$$)$ The observed RR Lyrae
variables are from \citet{pri01,pri02,pri03}. Solid and open squares represent
{\it ab}-type and {\it c}-type RR Lyrae stars, respectively. Type $c$ RR Lyrae
were fundamentalized by Log $P_F$ = Log $P_c$ + 0.13
\citep{cas87,lee90}. In order to avoid small number statistics, the catalogs of
NGC 6388 (12 RR Lyraes from \citet{pri02}) and NGC 6441 (63 RR Lyraes from
\citet{pri01,pri03}) were combined. Dashed line in ($a$) denotes the observed
mean $P_F$ value ($\langle P_{F} \rangle$ = 0.661 day) for the combined NGC 6388
(magenta) and NGC 6441(cyan) data. Solid line is the least-squares fit to the
data. Panel ($d$) is identical to panel ($a$). $($$b$, $c$, $e$, \& $f$$)$
Examples of the Monte Carlo simulations. Solid lines are the same as in ($a$ \&
$d$). Dashed and dotted lines in ($b$ \& $e$) and ($c$ \& $f$) denote the mean
values of $P_F$ and their 1 $\sigma$ uncertainties of 10,000 synthetic
simulations for the AMR model and the SHR model, respectively. \label{fig4}}
\end{center}
\end{figure*}

\clearpage
\begin{center}
\includegraphics[width=14cm]{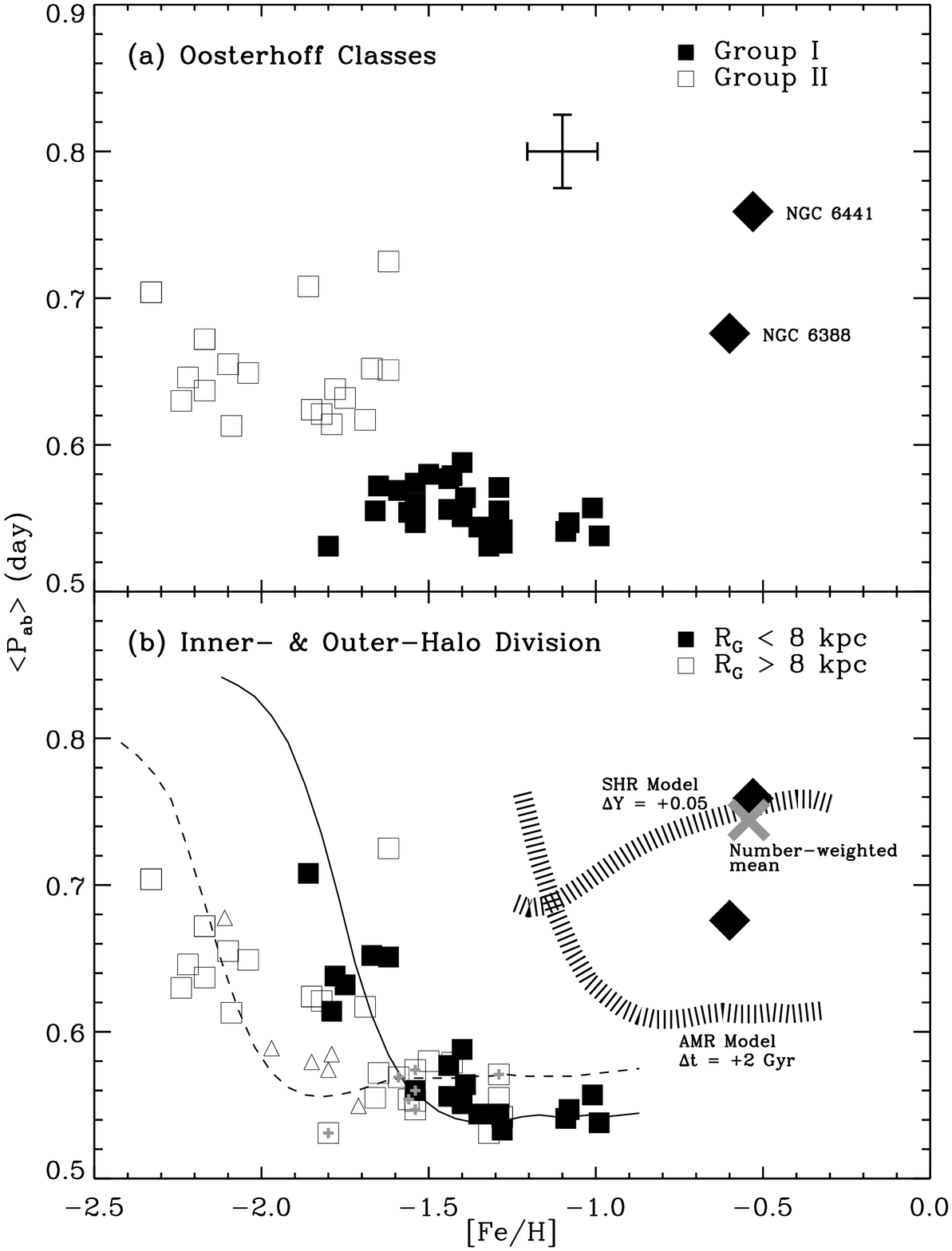}
\end{center}

\begin{figure*}
\begin{center}
\caption{ ($a$) Oosterhoff classes. The correlation between the mean period of
type {\it ab} RR Lyraes ($\langle P_{ab} \rangle$) and [Fe/H] for GCs in the
Galaxy is shown. The data are from \citet{cle01}. GCs having 5 or
more type {\it ab} RR Lyraes are shown. The error bar denotes the mean errors.
GCs are mainly divided into two distinct groups according to the mean period
and the metal abundance: Oosterhoff group I (filled squares; $\langle P_{ab}
\rangle$ $\simeq$ 0.55 days, and [Fe/H] $>$ $-$1.8) or group II (open squares;
$\langle P_{ab} \rangle$ $\simeq$ 0.65 days, and [Fe/H] $<$ $-$1.6). The
$\langle P_{ab} \rangle$ values for NGC 6388 (0.676 days from nine type-$ab$
variables) and for NGC 6441 (0.759 days from 42 type-$ab$ variables) are
obtained from \citet{cor06}. Note that NGC 6388 and NGC 6441 (solid diamonds)
belong to neither Oosterhoff group. 
($b$) Inner- \& outer-halo division. 
Solid and dashed lines are our model predictions for the 13-Gyr and
11.8-Gyr populations, respectively. GCs with the Galactocentric distance, $R_G$
$<$ 8 kpc (filled squares) are well reproduced by the models for the old
population (solid line), whereas most GCs with $R_G$ $>$ 8 kpc (open squares)
follow the model locus for slightly younger ages (dashed line). Small plus signs
mark GCs on retrograde orbits around the Galactic center, whereas triangles
represent the old GCs found in the LMC halo. Note that the retrograding GCs and the LMC
GCs follow the dashed line. Large gray cross between the diamonds marks
the number-weighted mean values of [Fe/H] and $\langle P_{ab} \rangle$ for NGC 6388
and NGC 6441.
Two thick stripes indicate the AMR model ($\Delta$t = +2 Gyr,
lower stripe) and the SHR model ($\Delta$Y = +0.05, upper stripe).  Note that the
SHR model prediction succeeds in reproducing the unusual positions of NGC 6388
and NGC 6441. \label{fig5}}
\end{center}
\end{figure*}

\end{document}